\documentclass{article}
\usepackage{graphicx} 
\usepackage{amsmath} 

\title{\vspace{-2cm}\bfseries\Large 
The Gauge Principle and Foundations of the Standard Model:\\[0.3em]
A Pedagogical Introduction Through QED}

\author{\normalsize Taha Anwar\\[0.25em]
\normalsize Department of Physics, Quaid-i-Azam University, Islamabad\\[0.25em]
\normalsize \texttt{tahaanwar607@gmail.com}}

\date{\vspace{0.5em}\normalsize August 2025}

\begin{document}

\maketitle
\vspace{3em}
\begin{abstract}
The Standard Model of particle physics is founded on the principle of local gauge symmetry and stands as one of the greatest achievements of modern science. Yet, its Lagrangian presents a steep learning curve, often introduced in contexts that assume extensive prior training. This review serves as a pedagogical bridge for advanced undergraduates. Using quantum electrodynamics (QED) as the simplest example, we illustrate how the gauge principle shapes the structure of the Standard Model Lagrangian, and connect familiar physics to modern field theory. Starting from the Lagrangian formulation of classical mechanics, special relativity, and basic quantum mechanics, we develop classical field theory, reformulate electromagnetism as a relativistic field theory, introduce the Dirac equation for spin-$\tfrac{1}{2}$ particles, and finally arrive at the idea of gauge symmetry. By showing how the demand for local \(U(1)\) invariance introduces the electromagnetic interaction, we highlight the essence of what it means for the Standard Model to be a gauge theory.

\end{abstract}
\newpage
\tableofcontents
\newpage
\section{Introduction}
The Standard Model of particle physics is one of the greatest achievements of modern science. It provides a remarkably accurate description of the electromagnetic, weak, and strong interactions in a single framework, based on the principle of local gauge invariance. Despite its success, students encountering the Standard Model for the first time often face a steep learning curve: the theory is usually presented in a formal language that assumes familiarity with classical field theory and quantum field theory, leaving the conceptual logic that links the undergraduate physics to the Standard Model.

The purpose of this review is pedagogical. We wish to illustrate the essential logic behind the construction of the Standard Model Lagrangian for undergraduates by developing its simplest component: the gauge structure of quantum electrodynamics (QED). 

Starting from Lagrangian formulation of classical mechanics, special relativity, and basic quantum mechanics, we develop classical field theory, reformulate electromagnetism as a relativistic field theory, introduce the Dirac equation for spin-$\tfrac{1}{2}$ particles, and finally arrive at the idea of gauge symmetry.

By showing how the demand for \emph{local U(1)} invariance necessarily introduces the electromagnetic interaction, we highlight the essence of what it means for the Standard Model to be a gauge theory.

The paper is organized to make the logical progression explicit. In Section~2 we review the Lagrangian formulation for continuous systems and introduce classical field theory with emphasis on intuition and examples. Section~3 presents relativistic electromagnetism in covariant form. Section~4 introduces the Lagrangian formalism of electromagnetism and demonstrates what the Lagrangian of electrodynamics looks like. 

Section~5 introduces the relativistic matter description and highlights global phase symmetry. While section~6 demonstrates how the Lagrangian of electrodynamics (Introduced in Section~4) arises naturally by promoting global \(U(1)\) symmetry to a local \(U(1)\)  symmetry. We close with a short concluding remark that places the QED construction in the broader context of the Standard Model and points to how the same symmetry-based logic extends to other gauge groups. 

\section{Classical Field Theory}
\subsection{Lagrangian Mechanics}
In classical mechanics, the motion of a point particle is obtained from
the principle of stationary action. We define the action as the time
integral of the Lagrangian.
\begin{equation}
    S[q] = \int_{t_i}^{t_f} L\!\left(q(t), \dot{q}(t), t\right) \, dt ,
\end{equation}
where $q(t)$ denotes the coordinate of the particle and $\dot{q}(t)$ denotes its velocity. To find the actual trajectory, we imagine all
possible paths $q(t)$ connecting the fixed endpoints
\[
    q(t_i) = q_i, 
    \qquad 
    q(t_f) = q_f,
\]
and require that the action be stationary under small variations
$\delta q(t)$ of the path. This yields the Euler--Lagrange equation
\begin{equation}\label{EL_particle}
    \frac{d}{dt} \left( \frac{\partial L}{\partial \dot{q}_i} \right) - \frac{\partial L}{\partial q_i} = 0,
\end{equation}
which is the equation of motion for the particle.

\subsection{Extending to fields}
In field theory, the dynamical variable is not a single coordinate
$q(t)$, but a field $\phi(x,t)$ defined at each point in space and
time. The field has infinitely many degrees of freedom --- one at each
spatial point. The analogue of the particle trajectory is now a field
configuration evolving in time.

The action is obtained by integrating the Lagrangian density over 4 dimensions of spacetime.
\begin{equation}\label{field_lagrangian}
    S[\phi] = \int d^4x \, \mathcal{L}\!\left(\phi(x), \partial_\mu \phi(x)\right) ,
\end{equation}
where \(\partial_\mu = \frac{\partial}{\partial x^\mu}\) and 
\(\mathrm{d}^4x = \mathrm{d}^3x \, \mathrm{d}t\).
.

To find the equation of motion for $\phi$, we consider all possible
field configurations $\phi(x)$ that agree with given boundary values at
the initial and final times $t_i$ and $t_f$, but are otherwise arbitrary
in between. In analogy with mechanics, we require the action to be
stationary under variations $\delta\phi(x)$ of the field, i.e.
\[
    \delta S = 0 \qquad \text{with} \qquad 
    \delta\phi(x,t_i) = \delta\phi(x,t_f) = 0.\]
Carrying out this variation gives the field-theoretic Euler--Lagrange
equation,
\begin{equation}\label{field_EL_eq}
    \frac{\partial \mathcal{L}}{\partial \phi}
    - \partial_\mu \left( \frac{\partial \mathcal{L}}{\partial (\partial_\mu \phi)} \right) = 0
\end{equation}

This is the general equation of motion for a classical field. In
practice, one specifies a particular Lagrangian density for the
system under study, substitutes it into Eq.~\eqref{field_EL_eq}, and
obtains the equations of motion for the field $\phi$.
\subsection*{Example: Motion of a One-Dimensional String}

As an illustrative example of classical field theory, consider the transverse displacement of a one-dimensional string, described by a scalar field \(\phi(x,t)\). In analogy with classical mechanics, we construct the Lagrangian density as the difference between kinetic and potential energy densities.

The kinetic energy density comes from the time derivative of the field, while the potential energy density arises from the spatial gradient (which represents the elastic restoring force of the string). Thus, the Lagrangian density is
\begin{equation}\label{string_lagrangian}
     \mathcal{L} = \frac{1}{2}\left( \dot{\phi}^2 - (\phi')^2 \right)
\end{equation}
where \(\dot{\phi} \equiv \partial_t \phi\) and \(\phi' \equiv \partial_x \phi\).
Using \eqref{string_lagrangian} in the Euler lagrange equation for fields \eqref{field_EL_eq}
\[
\frac{\partial \mathcal{L}}{\partial \phi} = 0, 
\quad 
\frac{\partial \mathcal{L}}{\partial (\partial_t \phi)} = \dot{\phi}, 
\quad 
\frac{\partial \mathcal{L}}{\partial (\partial_x \phi)} = -\phi'.
\]
And taking derivatives gives
\[
\partial_t \left( \dot{\phi} \right) + \partial_x \left( -\phi' \right) = 0,
\]
or equivalently,
\begin{equation}
\frac{\partial^2 \phi}{\partial t^2} - \frac{\partial^2 \phi}{\partial x^2} = 0
\label{eq:wave_eq}
\end{equation}
which is the familiar one-dimensional wave equation. This shows how the field-theoretic formalism naturally reproduces the dynamics of a vibrating string.
\subsection{Plane-wave solutions and dispersion relation}

Consider a plane-wave ansatz for \eqref{eq:wave_eq}
\begin{equation}
\phi(x,t) = A\,e^{i(kx - \omega t)} ,
\label{eq:ansatz}
\end{equation}
with constant amplitude $A$, wave number $k$, and angular frequency $\omega$.
Substituting \eqref{eq:ansatz} into \eqref{eq:wave_eq} gives
\begin{equation}
(-\omega^2 + k^2)\,A\,e^{i(kx - \omega t)} = 0
\quad \Longrightarrow \quad
\omega^2 = k^2 .
\label{eq:dispersion}
\end{equation}
Thus the dispersion relation is
\[
\omega = \pm\, |k| .
\]
Both the phase and group velocities are
\[
v_{\mathrm{ph}} \;=\; \frac{\omega}{k} \;=\; \pm 1,
\qquad
v_{\mathrm{g}} \;=\; \frac{d\omega}{dk} \;=\; \pm 1 \,.
\]
which reflects a nondispersive medium in natural units ($c=1$). Identifying $E=\hbar\omega$ and $p=\hbar k$, Eq.~\eqref{eq:dispersion} implies
\begin{equation}
E^2 = p^2 ,
\label{eq:massless_relation}
\end{equation}
which is the relativistic energy--momentum relation for a massless particle. 
\subsection{Adding a quadratic potential: massive field}

If we include a quadratic potential in the Lagrangian density,
\[
\mathcal{L} \;=\; \frac{1}{2}(\partial_t \phi)^2 \;-\; \frac{1}{2}(\partial_x \phi)^2 \;-\; \frac{1}{2} m^2 \phi^2 .
\label{eq:L_massive_1d}
\]
and apply the Euler Lagrange equation for fields, we
get the Klein--Gordon equation in $1+1$ dimensions:
\begin{equation}
\partial_t^2 \phi \;-\; \partial_x^2 \phi \;+\; m^2 \phi \;=\; 0 .
\label{eq:KG_1d}
\end{equation}
If we again start with a plain wave ansatz \eqref{eq:ansatz},
substitution into \eqref{eq:KG_1d} yields
\begin{equation}
(-\omega^2 + k^2 + m^2)\,A\,e^{i(kx - \omega t)} = 0
\quad \Longrightarrow \quad
\omega^2 \;=\; k^2 + m^2 .
\label{eq:disp_massive}
\end{equation}
Identifying energy and momentum via \(E=\omega\) and \(p=k\) (in natural units), we see that
\eqref{eq:disp_massive} is the relativistic energy--momentum relation for a massive particle:
\begin{equation}
E^2 \;=\; p^2 + m^2 .
\end{equation}
The group velocity satisfies
\[
v_g \;=\; \frac{d\omega}{dk} \;=\; \frac{k}{\sqrt{k^2 + m^2}} \;<\; 1 \,.
\]
as expected for massive excitations.

\paragraph{Remark.}
In field-theory language, the term \( \tfrac12 m^2\phi^2 \) is the \emph{mass term} coming from
a field potential \(V(\phi)=\tfrac12 m^2\phi^2\).
The massless case corresponds to \(V(\phi)=0\), i.e.\ the free field
\(\mathcal{L}=\tfrac12(\partial_t\phi)^2-\tfrac12(\partial_x\phi)^2\),
whose dispersion is \(\omega^2=k^2\).
\subsection{Lorentz covariance}

A central postulate of Special Relativity is that the laws of physics are the same in all inertial frames. This means that the equations describing these laws should retain the same form regardless of the inertial frame of reference, a property called Lorentz covariance. In field theory, this is ensured by starting from Lagrangians, that are Lorentz invariant, i.e. constructed from scalar quantities formed by contracting four-vectors or tensors, so that the resulting equations of motion automatically have the same form in every inertial frame.

Below we illustrate this for the same real scalar field theory as in previous section. We use Einstein summation convention: repeated upper and lower indices are summed. Our metric convention is
\[
\eta_{\mu\nu}=\mathrm{diag}(+1,-1,-1,-1),
\]
so that for any two four-vectors $a^\mu,b^\mu$ the Minkowski inner product is $a\cdot b = \eta_{\mu\nu}a^\mu b^\nu$. The four gradient is $\partial_\mu = \frac{\partial}{\partial x^\mu}$ and the raised index version is
\[
\partial^\mu = \eta^{\mu\nu}\partial_\nu.
\]
The Lorentz invariant second order operator (the d'Alembertian) is
\[
\partial_\mu\partial^\mu \,=\, \eta^{\mu\nu}\partial_\mu\partial_\nu \,=\, \frac{\partial^2}{\partial t^2} - \nabla^2.
\]
Because, this is built from contracted indices it is invariant under Lorentz transformations: its numerical value at a spacetime point is the same (when evaluated on objects with the appropriate transformation properties).

A minimal Lorentz invariant Lagrangian density for a free real scalar field $\phi(x^\mu)$ is
\begin{equation}
\mathcal{L} \;=\; \tfrac{1}{2}\,\partial_\mu \phi\,\partial^\mu \phi
\end{equation}
To derive the equation of motion use the Euler\--Lagrange equation for fields
\[ \frac{\partial \mathcal{L}}{\partial \phi} = 0\]

\[\frac{\partial \mathcal{L}}{\partial(\partial_\mu\phi)} = \partial^\mu\phi \]
Hence
\begin{equation}
\partial_\mu \partial^\mu \, \phi(x^\mu) \;=\; 0 \,
\end{equation}
which is the same \textbf{massless Klein\--Gordon equation} as in previous section. Because it was obtained from a manifestly Lorentz invariant Lagrangian built from contracted indices, the equation has the same form in every inertial frame, since the two 4-vectors with upper and lower indices transform in the opposite ways under a Lorentz transformation.

\subsubsection{Lorentz covariance vs Lorentz invariance (short clarification)}
\begin{itemize}
\item A \emph{Lorentz scalar} is a quantity whose numerical value is unchanged under Lorentz transformations (e.g. $\partial_\mu\phi\,\partial^\mu\phi$).
\item A field equation is called \emph{Lorentz covariant} if it transforms consistently under Lorentz transformations, so that it's form is the same in all inertial frames. For example, $\partial_\mu \partial^\mu \, \phi(x^\mu) \;=\; 0 \,$ is covariant.
\item A Lagrangian density is \emph{Lorentz invariant} if it is a scalar under Lorentz transformations. Invariance of the action guarantees covariance of the equations of motion.
\end{itemize}

\vspace{1ex}

\textbf{Takeaway:} Build Lagrangians from combinations of Lorentz scalars (contracted indices). Euler\--Lagrange equations derived from Lorentz invariant Lagrangians are Lorentz covariant and therefore describe dynamics whose \emph{form} is identical in all inertial frames.

\section{Relativistic Electromagnetism}

As discussed in the previous section, the construction of a Lorentz covariant field theory requires quantities that transform appropriately under Lorentz transformations. In particular, we must build the theory from Lorentz invariant or covariant objects. For electromagnetism, the first step is to introduce the four-current.

\subsection{The Four-Current}

In non-relativistic physics, one usually begins with the charge density $\rho(x,t)$ as the fundamental quantity describing the distribution of charge in space. However, charge density by itself does not transform consistently between different inertial frames. This is because the volume element changes under Lorentz transformations due to length contraction, so $\rho$ is frame-dependent.

A better perspective arises when we reinterpret charge density as a flux through time. Since
\[
\rho = \frac{\Delta Q}{\Delta x \, \Delta y \, \Delta z} \,
\]
we may think of it as the amount of charge crossing a unit volume in the direction of time axis. In a different frame, part of this flux will appear as charge density, while part of it will appear as spatial current density $\vec{j}$, which measures the flow of charge through a unit surface per unit time in a particular spatial direction:
\[
j_z = \frac{\Delta Q}{\Delta t \, \Delta x \, \Delta y} \,
\]
Thus, when space and time are treated on equal footing as in relativity, the natural object is the \emph{four-current}, which unifies charge density and current density:
\begin{equation}
j^\mu \;=\; (c\,\rho, \vec{j}) \,
\end{equation}

Here, $j^0 = c \, \rho$ is the time component, while $j^i$ $(i=1,2,3)$ represent the spatial components of the current density, where $c$ is introduced for dimensional consistency. This object transforms as a four-vector under Lorentz transformations, and can be used to build a Lorentz scalar.
\subsection{Electromagnetic 4-potential}

Motivated by the scalar and vector potentials of electrostatics and magnetostatics, we can define an electromagnetic potential from the 4-current as following
\[
A^\mu \;=\; \frac{\mu_0}{4\pi}\int \frac{j^\mu}{|\vec{r}-\vec{r}'|}\, d^3r'
\]
Using \(c^2=\dfrac{1}{\mu_0\varepsilon_0}\) (so \(\mu_0 c = \sqrt{\mu_0/\varepsilon_0} = 1/(\varepsilon_0 c)\)), we rewrite zeroth component of the above expression as
\[
A^0 \;=\; \frac{\mu_0}{4\pi}\int \frac{c\,\rho}{|\vec{r}-\vec{r}'|}\, d^3r'
\;=\; \frac{1}{4\pi\varepsilon_0\,c}\int \frac{\rho}{|\vec{r}-\vec{r}'|}\, d^3r' 
\]
This shows that the zeroth component of this electromagnetic potential is actually just the electrostatic potential \(\phi\) 
\[
A^0 \;=\; \frac{\phi}{c}.
\]
Meanwhile, the spatial components of the 4-vector $A$ are given by
\[
A^i \;=\; \frac{\mu_0}{4\pi}\int \frac{j^i}{|\vec{r}-\vec{r}'|}\, d^3r'
\]
Which is actually just the magnetic vector potential 
Thus, the electromagnetic 4-potential takes the compact form
\begin{equation}
A^\mu \;=\; \left(\frac{\phi}{c},\, \vec{A}\right)
\end{equation}
\subsection{Electromagnetic Field Tensor}

To motivate the unification of electric and magnetic fields, consider a simple physical setup. Suppose we have a conventional current flowing in a long, straight wire, and a positive test charge moving parallel to the wire in the same direction as the current.  

In the rest frame of the wire, the test charge experiences a magnetic force directed radially inward toward the wire. However, in the rest frame of the moving charge, the situation appears different. In this frame, the positive charges responsible for the conventional current move at a lower velocity, while the negative charges (the electrons) move in the opposite direction at a higher velocity. Due to special relativity, the higher velocity of the negative charges leads to a greater length contraction, producing a net negative charge density in the wire. As a result, in this frame the test charge experiences an attractive electric force toward the wire.  

Thus, what appears as a purely magnetic interaction in one frame manifests as a purely electric interaction in another. This illustrates that electric and magnetic fields are not independent entities, but instead are two aspects of a single physical object that transforms consistently under Lorentz transformations. This unified description is achieved through a second-rank antisymmetric tensor, known as the \emph{electromagnetic field tensor} (or Faraday tensor), which combines both electric and magnetic fields in a covariant framework. (For a detailed discussion, see Chapter~12 of \cite{griffiths2017}.)

\[
F^{\mu\nu} \;=\;
\begin{pmatrix}
0 & -E_x/c & -E_y/c & -E_z/c \\[6pt]
E_x/c & 0 & -B_z & B_y \\[6pt]
E_y/c & B_z & 0 & -B_x \\[6pt]
E_z/c & -B_y & B_x & 0
\end{pmatrix}
\]
This is the matrix representation of the tensor, which you can build by unifying electric and magnetic fields in a way, such that they transform nicely under a Lorentz transformation $x'^\mu=\Lambda^\mu{}_{\!\nu}x^\nu$ as the following
\[
F'^{\mu\nu}(x') \;=\; \Lambda^\mu{}_{\!\alpha}\,\Lambda^\nu{}_{\!\beta}\;F^{\alpha\beta}(x),
\]
Another elegant feature of the Faraday tensor is that it can be expressed directly in terms of the four-potential:
\begin{equation}\label{faraday_tensor}
F^{\mu\nu} \;=\; \partial^\mu A^\nu \;-\; \partial^\nu A^\mu
\end{equation}
This compact expression automatically guarantees the antisymmetry of $F^{\mu\nu}$ and encodes both electric and magnetic fields as derivatives of the four-potential. We can check the consistency between two descriptions of the Faraday tensor explicitly
\[
F^{i0} \;=\; \partial^i A^0 - \partial^0 A^i
\]
With our convention, we have $\partial_i=-\partial^i$, so the above expression becomes

\[ F^{i0} = \partial^i\left(\frac{\varphi}{c}\right) - \partial^0 A^i \nonumber\\
       = -\,\frac{1}{c}\,\partial_i \varphi \;-\; \frac{1}{c}\,\frac{\partial A^i}{\partial t} \nonumber\\
\]
This is precisely the standard relation (divided by speed of light)
\[
\mathbf{E} = -\nabla \varphi - \frac{\partial \mathbf{A}}{\partial t}
\]
From the matrix representation, we also know that $F^{i0} = \frac{1}{c}\,E^i$. 
This demonstrates the consistency between two definitions of the Faraday tensor for electric fields. A similar demonstration can be done for the magnetic field components as well, and has been left as an exercise for the reader.
\section{Electromagnetism as a Field Theory}

\subsection{Relativistic Free Particle}

We begin with the relativistic free particle before introducing interactions.  
The action must have dimensions of energy $\times$ time and must also respect Lorentz invariance.  
Both the rest energy $mc^2$ and the proper time element $d\tau$ are Lorentz-invariant quantities.  
Thus, a natural choice for the action is
\begin{equation}
    S = -mc^2 \int d\tau,
\end{equation}
Here, minus sign is used as a convention. Writing the proper time in terms of coordinate time $t$,
\[
    d\tau = dt \, \sqrt{1 - \frac{v^2}{c^2}},
\]
we obtain the Lagrangian
\[
    L = -mc^2 \sqrt{1 - \frac{v^2}{c^2}}
\]
The canonical momentum is defined as
\[  p_i = \frac{\partial L}{\partial v^i}.
\]

\begin{align}
    p_i &= -mc^2 \, \frac{1}{2}\left(1 - \frac{v^2}{c^2}\right)^{-1/2}\left(-\frac{2 v_i}{c^2}\right) \nonumber \\
        &= \frac{mv_i}{\sqrt{1 - \tfrac{v^2}{c^2}}}.
\end{align}

If we raise the index $i$, this is precisely the relativistic momentum 
\[
    \mathbf{p} = \gamma m \mathbf{v}, \quad \gamma = \frac{1}{\sqrt{1 - v^2/c^2}}.
\]
This shows that the familiar relativistic momentum naturally arises from the canonical formalism applied to the Lorentz-invariant free particle Lagrangian.
\subsection{Dynamics of a charged particle in an EM field}
Dynamics of a charged particle in an electromagnetic field depends on it's interaction with the field at it's location in space and time encoded by the potential energy. A simple way to include this interaction in the action for dynamics of these charged particles, which respects Lorentz invariance is to use the invariant product of space-time displacement and the 4-potential at the location.
\begin{equation}\label{eq:action_charged}
S \;=\; -m c^2 \int d\tau \;-\; q\int A_\mu(x)\,dx^\mu 
\end{equation}
Writing \(dx^\mu=(c\,dt,\,d\mathbf{x})\) and noting
\[
A_\mu\,dx^\mu = A_0\,dx^0 + A_i\,dx^i = \frac{\varphi}{c}\,c\,dt - \mathbf{A}\cdot d\mathbf{x}
= \varphi\,dt - \mathbf{A}\cdot d\mathbf{x},
\]
we may express the action as \(S=\int L\,dt\) with the Lagrangian
\[
L(\mathbf x,\mathbf v,t) \;=\; -mc^2\sqrt{1-\frac{v^2}{c^2}} \;-\; q\varphi(\mathbf x,t) \;+\; q\,\mathbf{A}(\mathbf x,t)\!\cdot\!\mathbf v,
\]
where \(\mathbf v=d\mathbf x/dt\).

This Lagrangian may be familiar to you from your course in Lagrangian mechanics, when relativistic energy is replaced by non-relativistic kinetic energy.
The Lagrangian under consideration can also be expressed as
\[
L \;=\; -mc^2 \sqrt{1 - \frac{v^2}{c^2}} \;+\; q\Big(- A_i v^i - A_0c \Big),
\]
We can derive the equation of motion for the particle by using the Euler Lagrange equation for particles \eqref{EL_particle}
\paragraph{L.H.S}
\[
\frac{\partial L}{\partial v^i} \;=\; \frac{m v_i}{\sqrt{1 - v^2/c^2}} \;-\; q A_i \;=\; \mathbf{p} - q A_i,
\]
While evaluating the total time derivative, we need to remember that vector potential for a particle can change for two reasons:

1) If the vector potential itself is time dependent. This is 
encoded by the partial derivative w.r.t time.

2) As the position of particle is changed, vector potential 
also changes for the particle. So, we have to use the chain 
rule.
  
\[
\frac{d}{dt}\left(\frac{\partial L}{\partial v^i}\right) \;=\; \frac{d\mathbf{p}}{dt} - q \frac{\partial A_i}{\partial t} - q \frac{\partial A_i}{\partial x^j}\frac{dx^j}{dt}
\]
\paragraph{R.H.S}
\[
\frac{\partial L}{\partial x^i} \;=\; q\left(-\frac{\partial A_j}{\partial x^i} v^j - \frac{\partial A_0c}{\partial x^i}\right)
\]
Thus, the Euler–Lagrange equation gives, after rearranging:
\[
\frac{d\mathbf{p}}{dt} \;=\; q\left(\frac{\partial A_i}{\partial t} - \frac{\partial A_0c}{\partial x^i} + \big(\frac{\partial A_i}{\partial x^j} -\frac{\partial A_j}{\partial x^i} \big) v^j\right)
\]
Here, we can recoginze the first term as electric field ($\vec{A}=A^i = -A_i$)
\[
\mathbf{E} = -\frac{\partial \mathbf{A}}{\partial t} - \frac{\partial \phi}{\partial x^i}
\]
While the second term can be recognized as the cross product of velocity and magnetic field 
\[
\Big(\frac{\partial A_i}{\partial x^j} - \frac{\partial A_j}{\partial x^i}\Big) v^j.
\]
After raising the index on $A$, the first term can be recognized as 
\(-(\vec{v}\cdot\nabla)\vec{A}\), while the second term is 
\(\nabla(\vec{v}\cdot\vec{A})\). Using the vector identity
\[
\nabla(\vec{v}\cdot\vec{A}) - (\vec{v}\cdot\nabla)\vec{A}
= \vec{v}\times(\nabla\times\vec{A}),
\]
And since \(\vec{B} = \nabla \times \vec{A}\), we identify
\[
\Big(\frac{\partial A_i}{\partial x^j} - \frac{\partial A_j}{\partial x^i}\Big) v^j
= \vec{v} \times \vec{B},
\]
We get the following E.O.M

\begin{equation}
\frac{d\mathbf{p}}{dt} \;=\; q\left( \mathbf{E} + \mathbf{v} \times \mathbf{B} \right)
\end{equation}
Thus, starting from the action \eqref{eq:action_charged}, we have recovered the Lorentz force law.

\textbf{Takeaway:} We started by adding an $interaction\,\ term$ to the lagrangian for a free particle that only encodes the motion of a particle by itself ($kinetic \,\ term$), used the Euler Lagrange equation, and derived the equation of motion that encodes how a charged particle is influenced by an electromagentic field. But the electromagentic field is also influenced by the presence of a charged particle and it's motion, and has a life of it's own.

\subsection{Dynamics of EM field in presence of charge and current distributions}

Evolution of a free field is encoded by how the field varies in spacetime, so the kinetic (propagation) part of a field Lagrangian is built from derivatives of the field. For the electromagnetic potential \(A^\mu(x)\), the natural derivative object to use is the antisymmetric Faraday tensor \eqref{faraday_tensor}.

To form a Lorentz scalar from $F^{\mu\nu}$, we must contract indices; the simplest nontrivial scalar built only from \(F^{\mu\nu}\) is \(F_{\mu\nu}F^{\mu\nu}\). Since the Lagrangian density for a field theory has units of energy density, and $F^{\mu\nu}F_{\mu\nu}$ has units of magnetic field squared, we can divide the kinetic term by magnetic permeability $\mu_0$ for consistency. For the interaction term, we have two 4-vectors (4-potential field itself and the 4-current), so a natural choice is to use their Lorentz invariant product.
\begin{equation}\label{eq:EM_Lag_density}
\mathcal{L} \;=\; -\frac{1}{4\mu_0}\,F_{\mu\nu}F^{\mu\nu} \;-\; J_\mu A^\mu,
\end{equation}
We want to derive the equations of motion by varying \(\mathcal{L}\) with respect to the field \(A_\rho\). The field Euler--Lagrange equation reads
\begin{equation}\label{eq:fieldEL}
\partial_\sigma\!\left(\frac{\partial\mathcal{L}}{\partial(\partial_\sigma A_\rho)}\right)
- \frac{\partial\mathcal{L}}{\partial A_\rho} \;=\; 0 .
\end{equation}

Compute the two pieces separately.

\paragraph{(i)}
Only the interaction term depends explicitly on \(A_\rho\), so
\[
\frac{\partial\mathcal{L}}{\partial A_\rho} \;=\; -J^\rho .
\]

\paragraph{(ii)} For \(\partial\mathcal{L}/\partial(\partial_\sigma A_\rho)\), we need the derivative of \(F_{\mu\nu}\) with respect to the derivative \(\partial_\sigma A_\rho\).
Because \(\partial_\mu A_\nu\) and \(\partial_\nu A_\mu\) are independent components, the only times the derivative
\(\partial(\partial_\mu A_\nu)/\partial(\partial_\sigma A_\rho)\) is nonzero and 1, is when the derivative index and the potential index match those on the object we differentiate. 
Concretely,
\[
\frac{\partial(\partial_\mu A_\nu)}{\partial(\partial_\sigma A_\rho)}
= \delta_\mu^{\ \sigma}\,\delta_\nu^{\ \rho},
\]
and 
\[
\frac{\partial(\partial_\nu A_\mu)}{\partial(\partial_\sigma A_\rho)}
= \delta_\nu^{\ \sigma}\,\delta_\mu^{\ \rho}.
\]
Subtracting these two contributions gives
\begin{equation}\label{derlowf}
    \frac{\partial F_{\mu\nu}}{\partial(\partial_\sigma A_\rho)}
= \delta_\mu^{\ \sigma}\,\delta_\nu^{\ \rho} \;-\; \delta_\nu^{\ \sigma}\,\delta_\mu^{\ \rho}
\end{equation}
similarly,
\begin{equation}\label{derupf}
    \frac{\partial F^{\mu\nu}}{\partial(\partial_\sigma A_\rho)}
= \delta^{\mu \sigma}\,\delta^{\nu \rho} \;-\; \delta^{\nu \sigma}\,\delta^{\mu \rho}
\end{equation}
\textbf{Now}

\[
\frac{\partial\mathcal{L}}{\partial(\partial_\sigma A_\rho)}
= -\frac{1}{4\mu_0}\,\frac{\partial\big(F_{\mu\nu}F^{\mu\nu}\big)}{\partial(\partial_\sigma A_\rho)}.
\]

\[
\frac{\partial\big(F_{\mu\nu}F^{\mu\nu}\big)}{\partial(\partial_\sigma A_\rho)}
= F^{\mu\nu}\,\frac{\partial F_{\mu\nu}}{\partial(\partial_\sigma A_\rho)}
+ F_{\mu\nu}\,\frac{\partial F^{\mu\nu}}{\partial(\partial_\sigma A_\rho)} .
\]
Using \eqref{derlowf} and \eqref{derupf}

\begin{align*}
F^{\mu\nu}\big(\delta_\mu^{\ \sigma}\delta_\nu^{\ \rho}-\delta_\nu^{\ \sigma}\delta_\mu^{\ \rho}\big)
&= F^{\sigma\rho}-F^{\rho\sigma}=2F^{\sigma\rho},\\[4pt]
F_{\mu\nu}\big(\delta^{\mu\sigma}\delta^{\nu\rho}-\delta^{\nu\sigma}\delta^{\mu\rho}\big)
&= F^{\sigma\rho}-F^{\rho\sigma}=2F^{\sigma\rho}.
\end{align*}
Adding the two contributions yields
\[
\frac{\partial\big(F_{\mu\nu}F^{\mu\nu}\big)}{\partial(\partial_\sigma A_\rho)}
= 2F^{\sigma\rho} + 2F^{\sigma\rho} = 4F^{\sigma\rho}
\]
Hence
\[
\frac{\partial\mathcal{L}}{\partial(\partial_\sigma A_\rho)}
= -\frac{1}{4\mu_0}\cdot 4F^{\sigma\rho} = -\frac{1}{\mu_0}F^{\sigma\rho}.
\]
\paragraph{Equation of motion:}
We can plug the above expressions in the Euler lagrange equation to get the following equation of motion
\begin{equation}\label{eq:Maxwell_Eq}
\boxed{\qquad \partial_\sigma F^{\sigma\rho} \;=\; \mu_0\, J^\rho \qquad}
\end{equation}
Which are just the two non-homogeneous Maxwell's equations.
For the two homogeneous Maxwell's equations, we recall the definition of the field strength tensor
\[
F_{\mu\nu} = \partial_\mu A_\nu - \partial_\nu A_\mu .
\]
From this, consider the combination
\[
\partial_\lambda F_{\mu\nu} + \partial_\mu F_{\nu\lambda} + \partial_\nu F_{\lambda\mu}.
\]
Plugging in the definition, each term becomes a second derivative of the potential. For example,
\[
\partial_\lambda F_{\mu\nu} = \partial_\lambda \partial_\mu A_\nu - \partial_\lambda \partial_\nu A_\mu ,
\]
and similarly for the other terms. When we add the three cyclic permutations, every derivative cancels out because partial derivatives commute 
(\(\partial_\lambda \partial_\mu = \partial_\mu \partial_\lambda\)). Thus, we obtain the identity
\[
\partial_\lambda F_{\mu\nu} + \partial_\mu F_{\nu\lambda} + \partial_\nu F_{\lambda\mu} = 0 .
\]
This is called the \emph{Bianchi identity}, and it holds automatically for any \(F_{\mu\nu}\) defined in terms of a potential.
When we translate this compact tensor equation into vector form, it gives the two homogeneous Maxwell equations:
\[
\nabla \cdot \mathbf{B} = 0, 
\qquad 
\nabla \times \mathbf{E} + \frac{\partial \mathbf{B}}{\partial t} = 0 .
\]
So these two equations do not need to be postulated separately, they follow directly and inevitably from the definition of the Faraday tensor as the curl of the 4-potential.
\subsection{Lagrangian of Electrodynamics}
We have now derived both the Lorentz force law by varying the action with respect 
to the particle’s degrees of freedom, and Maxwell’s equations by varying with respect 
to the field. This illustrates the general spirit of a field theory in its Lagrangian 
formalism. 

Suppose we have two fields, $\phi$ and $\psi$, that interact with each other. A single, 
unified action for this theory typically contains three essential ingredients: 
\begin{enumerate}
    \item a kinetic term $\mathcal{L}_{\text{kin}}(\partial \phi)$ for $\phi$, which 
    describes how $\phi$ would evolve if it existed by itself,
    \item a kinetic term $\mathcal{L}_{\text{kin}}(\partial \psi)$ for $\psi$, which 
    similarly encodes the dynamics of $\psi$ alone,
    \item an interaction term $\mathcal{L}_{\text{int}}(\phi \, \psi)$, which couples 
    the two fields.
\end{enumerate}

Varying the action with respect to $\phi$, the kinetic term $\mathcal{L}_{\text{kin}}(\partial \phi)$ contributes second 
derivatives of $\phi$, while the interaction term contributes a source term 
proportional to $\psi$. This produces an equation of motion that describes how 
$\phi$ evolves in the presence of $\psi$. Similarly, varying with respect to $\psi$ 
gives an equation of motion for $\psi$ that is influenced by $\phi$. 

This essence of interacting field theories is directly reflected in our discussion 
of Electromagnetism, if we can show that dynamics of both charged matter and EM fields emerge coherently from a single Lagrangian framework. To demonstrate this, we need to show that the interaction term in the particle's action \eqref{eq:action_charged} is actually the same interaction term in the action for Maxwell's equations \eqref{eq:EM_Lag_density} (in the special case of a point particle).
For a point charge $q$ moving along the trajectory $\vec{x}(t)$, the $4$--current is written as
\[
J^\mu(x)\;=\; 
\big( c\rho,\, \vec{J} \big) 
= \Big( q c \, \delta^{(3)}\!\big(\vec{x}-\vec{x}(t)\big), \; q\vec{v}\, \delta^{(3)}\!\big(\vec{x}-\vec{x}(t)\big)\Big).
\]
Multiplying and dividing the zeroth component by $dt$, and using the chain rule $dx^\mu/dt = U^\mu \, (d\tau/dt)$, we can rewrite
\[
J^\mu(x) \;=\; q \, U^\mu \, \frac{d\tau}{dt} \; \delta^{(3)}\!\big(\vec{x}-\vec{x}(t)\big).
\]
Since a point particle should only couple to the potential at its own position $x(\tau)$, and since the current should be expressed in terms of Lorentz--invariant quantities like the proper time $\tau$, we insert an identity
\[
f(t) \;=\; \int d\tau \; f(\tau)\, \delta\!\big(t - t(\tau)\big).
\]
Using this in the current, we obtain
\[
J^\mu(x) \;=\; \int d\tau \; q \, U^\mu(\tau)\, 
\delta^{(4)}\!\big(x - x(\tau)\big).
\]
Plugging this into the interaction term of the Maxwell action
\[
S_{\text{int}} \;=\; - \int d^4x \; J^\mu(x)\, A_\mu(x),
\]
we find
\[
S_{\text{int}} \;=\; - \int d^4x \int d\tau \; q U^\mu(\tau) 
\delta^{(4)}\!\big(x - x(\tau)\big) A_\mu(x).
\]
The spacetime delta function enforces $x = x(\tau)$, leaving
\[
S_{\text{int}} \;=\; - q \int d\tau \; U^\mu(\tau) A_\mu\big(x(\tau)\big).
\]
Using the definition of 4-velocity, this is equivalently
\[
S_{\text{int}} = -q \int dx^\mu \, A_\mu.
\]
Which is exactly the interaction term for a relativistic charged particle coupled to the electromagnetic field. 

Hence, If there is one thing I want you to take away from this, it is that the Lagrangian of Electrodynamics has the elegant structure:

\begin{equation}\label{eq: ED_Lagrangian}
\boxed{\qquad \mathcal{L} \;=\; \mathcal{L}_{\text{kinetic for matter}}
\;-\; J^{\mu}(x) A_{\mu}(x)
\;-\; \frac{1}{4\mu_0} \, F^{\mu\nu} F_{\mu\nu} \qquad}
\end{equation}
\section{Quantum Mechanics}
\subsection{Non-relativistic Quantum Mechanics}
In physics, our central concern is to predict how a physical system will evolve in the future. The mathematical framework of any theory provides us with objects that describe the state of a system, and equations that govern their time evolution. For example, in classical mechanics the state of a particle is specified by its position and momentum, and their evolution is governed by Newton’s second law, $F=ma$.

In non-relativistic quantum mechanics, the state of a system is described by a wavefunction $\psi(\mathbf{x},t)$ or more generally a state vector $|\psi(t)\rangle$. Its time evolution is governed by the Schrödinger equation, which in essence is the Hamiltonian acting on the state:
\[
i\hbar \frac{\partial}{\partial t} \psi(\mathbf{x},t) = \hat{H} \psi(\mathbf{x},t).
\]
For a single particle of mass $m$ in a potential $V(\mathbf{x})$, the Hamiltonian is
\[
\hat{H} = \frac{\hat{\mathbf{p}}^2}{2m} + V(\mathbf{x}), \qquad 
\hat{\mathbf{p}} = -i\hbar \nabla, \quad \hat{E} = i\hbar \frac{\partial}{\partial t}.
\]
Thus the Schrödinger equation reads
\[
i\hbar \frac{\partial}{\partial t} \psi(\mathbf{x},t) = \left(-\frac{\hbar^2}{2m}\nabla^2 + V(\mathbf{x})\right)\psi(\mathbf{x},t).
\]
An important feature of the wavefunction in quantum mechanics is its 
\textbf{probabilistic interpretation}.  
The modulus squared gives the probability density,
\[
\rho(\mathbf{x},t) = |\psi(\mathbf{x},t)|^2,
\]
and integrating over all space gives the total probability of finding the particle:
\[
\int d^3x\, |\psi(\mathbf{x},t)|^2 = 1.
\]
This condition means that the \emph{state vector} $|\psi\rangle$ must always have unit norm.  
Therefore, any transformation that we apply to the state must preserve this normalization, 
otherwise the probability interpretation would break down.  

Mathematically, preserving the norm requires that the operator $U$ acting on $|\psi\rangle$ satisfies
\[
\langle \psi' | \psi' \rangle \;=\; \langle \psi | U^\dagger U | \psi \rangle \;=\; \langle \psi | \psi \rangle.
\]
Since this must hold for all $|\psi\rangle$, we conclude that
\[
U^\dagger U = 1,
\]
which is exactly the condition that $U$ be a \textbf{unitary operator}.  

A very simple example is multiplying the wavefunction by a complex phase factor:
\[
\psi(\mathbf{x},t) \;\longrightarrow\; \psi'(\mathbf{x},t) = e^{i\theta}\,\psi(\mathbf{x},t).
\]
Because $|e^{i\theta}|=1$, the probability density is unchanged:
\[
|\psi'|^2 = |e^{i\theta}\psi|^2 = |\psi|^2.
\]
Thus, the phase factor $e^{i\theta}$ acts as a unitary operator that preserves probabilities.  

Notice that this transformation is described by a single continuous parameter $\theta$.  
The set of all such phase transformations forms the group $U(1)$. 
Hence, the statement that physics is unchanged when we multiply a wavefunction by an overall phase 
is an example of a \textbf{$U(1)$ symmetry}. And It also has some association with the idea that probability is conserved. More on this later..

\subsection{Relativistic Quantum Mechanics}
 Schrodinger's equation is explicitly \textit{non-relativistic}. Time and space do not appear on equal footing, and the Hamiltonian is based on the non-relativistic kinetic energy. To incorporate special relativity, we must start from the relativistic energy-momentum relation
\[
E^2 = p^2 c^2 + m^2 c^4,
\]
and promote $E \to i\hbar \frac{\partial}{\partial t}$ and $\mathbf{p} \to -i\hbar \nabla$. This leads to the \textbf{Klein–Gordon equation}:
\[
\left( \frac{1}{c^2}\frac{\partial^2}{\partial t^2} - \nabla^2 + \frac{m^2 c^2}{\hbar^2} \right)\psi(\mathbf{x},t) = 0.
\]
This equation is relativistic and works well in QFT for spin-0 bosons. However, it is second order in time and it's solutions cannot be interpreted as probability density, unlike the wave function of Schrodinger's equation, because otherwise probability density would either not be conserved or could be negative.

Dirac recognized that the problem lay in the second-order structure of the equation. He proposed instead an equation \emph{first order} in both time and space derivatives — in effect, the ``square root’’ of the energy-momentum relation. 
He proposed the ansatz
\begin{equation}\label{linear_EM_relation}
    E = \boldsymbol{\alpha}\cdot \mathbf{p}\,c + \beta m c^2,
\end{equation}
where $\boldsymbol{\alpha}$ and $\beta$ are not just numbers, but some mathematical objects (matrices actually) which ensure that if we square both sides, we should recover the correct relativistic energy--momentum relation:
\[
E^2 = p^2c^2 + m^2 c^4 \]

This requirement puts restrictions on what kind of matrices $\boldsymbol{\alpha}$ and $\beta$ can be. For instance:  

\begin{itemize}
    \item When we square $\alpha_i p_i$, cross-terms like $\alpha_i \alpha_j p_i p_j$ appear. To reduce to just $p^2$, the matrices must be such that, these cross terms disappear.
    \item Concretely, this means that $\alpha_i$ and $\alpha_j$ must multiply in a way that behaves like $\delta_{ij}$.
    \item Similarly, $\alpha_i$ and $\beta$ must multiply in a way that cancels mixed terms, and $\beta^2$ must behave like the number $1$.
\end{itemize}

In more mathematical language, these conditions are written as
\[
\alpha_i \alpha_j + \alpha_j \alpha_i = 2\delta_{ij}, \qquad
\alpha_i \beta + \beta \alpha_i = 0, \qquad
\beta^2 = 1.
\]
You can check by squaring \eqref{linear_EM_relation}, and using above relations, that you recover the familiar relativistic energy momentum relation.

It turns out that the simplest way to satisfy all of these conditions is to use $4\times 4$ matrices.  
This means the wavefunction $\psi$ cannot be a single number, nor a two-component object, but must have four components.  
Such a four-component object is called a \textbf{Dirac bi-spinor}.  

With this setup, Dirac's equation for the wavefunction becomes
\[
i\hbar \frac{\partial}{\partial t}\psi = \left( -i\hbar c\, \boldsymbol{\alpha}\cdot\nabla + \beta m c^2 \right)\psi.
\]

We can rewrite this in a more compact way by introducing new symbols.  
Let us define
\[
\gamma^0 = \beta, \qquad \gamma^i = \alpha_i \quad (i=1,2,3).
\]
Using Einstein’s summation convention, where repeated indices are summed over, the equation above can be written as
\[
\big( i\hbar c\, \gamma^\mu \partial_\mu - mc^2 \big)\psi = 0.
\]

This is the celebrated \textbf{Dirac equation}, which is first-order in both space and time derivatives, fully relativistic, and naturally describes spin-$\tfrac{1}{2}$ particles like the electron.

Finally, as in any field theory, we can obtain this equation from a Lagrangian density. The \textbf{Dirac Lagrangian} is
\[
\mathcal{L}_{\text{Dirac}} = \bar{\psi}\left(i\hbar c\, \gamma^\mu \partial_\mu - mc^2\right)\psi,
\]
where $\bar{\psi} = \psi^\dagger \gamma^0$. Varying this Lagrangian with respect to $\bar{\psi}$ yields the Dirac equation. 

This Lagrangian forms the cornerstone of quantum field theory for fermions, and when coupled to the electromagnetic field it leads directly to quantum electrodynamics.

\section{Quantum Electrodynamics}
\subsection{Gauge invariance in electromagnetism}
Recall the definitions of electric and magnetic fields in terms of electric and magnetic vector potentials
\[
\mathbf{E} = -\nabla \phi - \frac{\partial \mathbf{A}}{\partial t}, 
\qquad
\mathbf{B} = \nabla \times \mathbf{A}.
\]

Now consider the following transformation of the potentials, where $\Lambda(\mathbf{x},t)$ is any smooth scalar function:
\[
\phi' = \phi + \frac{\partial \Lambda}{\partial t}, 
\qquad 
\mathbf{A}' = \mathbf{A} - \nabla \Lambda.
\]

If we compute the electric and magnetic fields from the new potentials, we find
\[
\mathbf{E}' = -\nabla \phi' - \frac{\partial \mathbf{A}'}{\partial t}, 
\qquad
\mathbf{B}' = \nabla \times \mathbf{A}'.
\]

Substituting $\phi'$ and $\mathbf{A}'$ shows that
\[
\mathbf{E}' = \mathbf{E}, \qquad \mathbf{B}' = \mathbf{B}.
\]

Thus, the physical fields $\mathbf{E}$ and $\mathbf{B}$ are unchanged, even though the potentials 
$(\phi, \mathbf{A})$ have been modified.  
This freedom to change the potentials without altering the observable physics is called 
\textbf{gauge invariance} in electromagnetism.
Now recall, that scalar and vector potentials can be combined into a single four-vector,
\[
A^\mu = \big( \phi/c, \, \mathbf{A} \big),
\]
In this notation, the gauge transformation can be written compactly as
\[
A^\mu \;\;\longrightarrow\;\; A'^\mu = A^\mu + \partial^\mu \Lambda,
\]
Which you can also write as follows, by lowering indices
\[
A_\mu \;\;\longrightarrow\;\; A'_\mu = A_\mu + \partial_\mu \Lambda,
\]
This demonstrates the idea that electromagnetic 4-potential is a field to which, if you add spacetime gradient of a scalar function, physics remains unchanged.
\subsection{U(1) Symmetry}
We momentarily set aside what we know about electromagnetism and imagine an empty universe.  
What basic principles should the laws of physics satisfy?

\begin{itemize}
    \item \textbf{Unitarity (probability conservation):} the norm of any quantum state must be preserved.
    \item \textbf{Lorentz invariance:} energy and momentum obey the relativistic relation
    \[
    E^2 = \mathbf{p}^2 c^2 + m^2 c^4.
    \]
\end{itemize}

These two requirements lead us to a Lagrangian for matter (fermions) in relativistic quantum field theory, the Dirac Lagrangian
\begin{equation}\label{dirac_lagrangian}
\mathcal{L}_D \;=\; \bar{\psi}\,\big(i\hbar c\,\gamma^\mu \partial_\mu - mc^2\big)\,\psi,
\end{equation}
which is Lorentz invariant and yields the Dirac equation via the Euler--Lagrange equations.

\subsubsection*{Global U(1) transformation}

Consider a \emph{global} phase change (the same phase everywhere in spacetime):
\[
\psi(x) \longrightarrow \psi'(x) = e^{i\theta}\,\psi(x), \qquad \bar{\psi}(x) \longrightarrow \bar{\psi}'(x) = \bar{\psi}(x)\,e^{-i\theta},
\]
with constant $\theta$. The phase cancels between $\psi$ and $\bar{\psi}$, and one checks directly that
\[
\mathcal{L}_D' \;=\; \bar{\psi}'\big(i\hbar c\,\gamma^\mu \partial_\mu - mc^2\big)\psi' \;=\; \mathcal{L}_D.
\]
Thus the Dirac Lagrangian has a global phase symmetry.

\subsubsection*{Local U(1) Transformation}

Starting with the Dirac lagrangian \eqref{dirac_lagrangian} again, now suppose we allow the phase to depend on spacetime:
\[
\psi \;\longrightarrow\; \psi' = e^{i\theta(x)}\,\psi,
\quad\quad
\bar{\psi} \;\longrightarrow\; \bar{\psi}' = e^{-i\theta(x)}\,\bar{\psi}.
\]
Let us substitute this into the Dirac Lagrangian. \\
\begin{enumerate}
    \item The mass term remains unchanged.
\[
\bar{\psi}'\,(-mc^2)\,\psi' 
= \bar{\psi}\,(-mc^2)\,\psi,
\]
    \item However, the derivative part transforms as following
\[
\bar{\psi}'\, i\hbar c\,\gamma^\mu \partial_\mu \psi'
= \bar{\psi}\, i\hbar c\,\gamma^\mu \partial_\mu \psi
\;-\; \hbar c\,\bar{\psi}\,\gamma^\mu (\partial_\mu \theta)\,\psi.
\]
\end{enumerate}
So, the transformed Lagrangian becomes
\[
\mathcal{L}_D' 
= \bar{\psi}\,\big( i\hbar c\,\gamma^\mu \partial_\mu - mc^2 \big)\,\psi
\;-\; \hbar c\,\bar{\psi}\,\gamma^\mu (\partial_\mu \theta)\,\psi.
\]
The extra piece proportional to $\partial_\mu \theta$ shows that
\[
\mathcal{L}_D' \;\neq\; \mathcal{L}_D.
\]
Therefore, the Dirac Lagrangian by itself is not invariant under a spacetime-dependent (local) phase transformation.

\subsubsection*{How to enforce local phase invariance}
Imagine performing some transformation of a lagrangian that consists of two terms $A$ and $B$.  
If under this transformation the two pieces change in the following way:
\[
A \;\longrightarrow\; A + C, 
\quad\quad
B \;\longrightarrow\; B - C,
\]
then the transformed Lagrangian becomes
\[
\mathcal{L}' = (A + C) + (B - C) = A + B = \mathcal{L}.
\]

So even though each term changes individually, their combined effect cancels out.  
The total Lagrangian remains the same before and after the transformation, meaning the theory is invariant under this transformation.

We can do something similar with the Dirac lagrangian. But first we note that the extra term $C$ is proportional to 4-gradient of a scalar function. So we need to introduce a new term in analogy with $B$, which contains a field that tranforms such that a 4-gradient of scalar function is added to it, and physics remain unchanged. As discussed in section 6.1, Electromagnetic 4-potential is exactly such a field. But keeping that aside, imagine a field $A$ which under local U(1) transformation of field $\psi$, transforms as follows

\[
A_\mu(x)\longrightarrow A'_\mu(x)=A_\mu(x)-\frac{\hbar c}{q}\,\partial_\mu\theta(x).
\]
Now, Start from the Dirac Lagrangian with a new term included:
\begin{equation}\label{incomplete_qed}
   \mathcal{L}
\;=\;
\bar{\psi}\big(i\hbar c\,\gamma^\mu\partial_\mu - mc^2\big)\psi
\;-\; q\,\bar{\psi}\gamma^\mu\psi\,A_\mu.
 \end{equation}
Perform the local U(1) transformation of this lagrangian and evaluate each piece. 
\begin{enumerate}
    \item The mass term is unchanged:
\[
\bar\psi'(-mc^2)\psi'=\bar\psi(-mc^2)\psi.
\]
    \item The derivative piece transforms as
\[
\begin{aligned}
\bar\psi' \, i\hbar c\,\gamma^\mu\partial_\mu\psi'
&= e^{-i\theta}\bar\psi \, i\hbar c\,\gamma^\mu\partial_\mu\big(e^{i\theta}\psi\big)\\[4pt]
&= \bar\psi\, i\hbar c\,\gamma^\mu\partial_\mu\psi \;-\; \hbar c\,\bar\psi\,\gamma^\mu(\partial_\mu\theta)\psi .
\end{aligned}
\]
    \item The new term becomes
\[
\begin{aligned}
- q\,\bar\psi'\gamma^\mu\psi'\,A'_\mu
&= -q\,\bar\psi\gamma^\mu\psi\;\Big(A_\mu - \frac{\hbar c}{q}\partial_\mu\theta\Big)\\[4pt]
&= -q\,\bar\psi\gamma^\mu\psi\,A_\mu \;+\; \hbar c\,\bar\psi\,\gamma^\mu(\partial_\mu\theta)\psi.
\end{aligned}
\]
\end{enumerate}
Adding the transformed derivative piece and the transformed new piece, the two extra terms proportional to \(\hbar c\,\bar\psi\gamma^\mu(\partial_\mu\theta)\psi\) cancel (one with plus sign, one with minus sign). Therefore every term in the transformed Lagrangian matches the corresponding term in the original Lagrangian, and we have
\[
 \mathcal{L}' = \mathcal{L} 
\]

Thus, we have achieved local U(1) symmetry by introducing a new field to the universe and adding a corresponding term to the Lagrangian. However, as discussed in section 4.4, the interpretation for this new term is actually that of an interaction term since it contains the product of two fields $\psi$ and $A$. The Lagrangian for field theory of two interacting fields should contain, in essence, three terms. An interaction term, and two kinetic terms which describe the kinetic evolution of two fields by themselves. But the Lagrangian \eqref{incomplete_qed}, does not allow the $A$ field to exist independently of $\psi$. The $A$ field should have a life of it's own i.e. we need to introduce a kinetic term for $A$ field, that must also maintain the local U(1) symmetry of our desired Lagrangian.\\
Since a kinetic term is built from derivatives of the potential, consider the derivative of the four-potential $\partial_\mu A_\nu$.

Under the gauge transformation
\[
A_\mu \longrightarrow A_\mu + \partial_\mu \Lambda,
\]
this derivative changes as
\[
\partial_\mu A_\nu \longrightarrow \partial_\mu A_\nu + \partial_\mu\partial_\nu \Lambda.
\]
Likewise,
\[
\partial_\nu A_\mu \longrightarrow \partial_\nu A_\mu + \partial_\nu\partial_\mu \Lambda.
\]
If we form the antisymmetric combination
\[
F_{\mu\nu} \equiv \partial_\mu A_\nu - \partial_\nu A_\mu,
\]
then, the extra second-derivative pieces cancel because partial derivatives commute. So, we can use this gauge invariant tensor to construct our desired kinetic term.\\
Finally, to ensure Lorentz invariance, we construct the scalar
\[
F^{\mu\nu} F_{\mu\nu},
\] 
and with the appropriate constant that ensures Lagrangian density has units of energy density (e.g. \( -\tfrac{1}{4\mu_0}\)) we end up obtaining the kinetic term of the electromagnetic field.

Adding this term to our \textit{incomplete} Lagrangian \eqref{incomplete_qed}, we obtain the Lagrangian for a theory of two interacting fields by imposing local $U(1)$ symmetry on the Dirac Lagrangian in a self-consistent manner.  

\begin{equation}\label{QEDlagrangian}
    \boxed{ \;
    \mathcal{L}
    = \bar{\psi}\big(i\hbar c\,\gamma^\mu \partial_\mu - mc^2\big)\psi
    \;-\; q\,\bar{\psi}\gamma^\mu\psi\,A_\mu
    \;-\; \frac{1}{4\mu_0}\,F_{\mu\nu}F^{\mu\nu}
    \;}
\end{equation}

\subsection{Lagrangian of QED}

In order to understand the essence of Lagrangian \eqref{QEDlagrangian}, we need to understand what the interaction term entails, specifically the object $q\,\bar{\psi}\gamma^\mu\psi$.

You may have seen Noether's theorem in a Lagrangian mechanics course: it states that
for every continuous symmetry of the Lagrangian there is an associated conserved quantity.
The same idea applies to field theories such as the Dirac field. For every symmetry of the Lagrangian, there is a conserved current. (For more details on this, see Section~4.5 of \cite{schwichtenberg2018}.)

We have seen that Dirac Lagrangian remains unchanged under a global U(1) or phase transformaion. Using Noether's theorem, the conserved current derived from global U(1) symmetry turns out to be $\bar\psi\gamma^\mu\psi$, which obeys the continuity equation:
\[\partial_\mu\big(\bar\psi\gamma^\mu\psi\big) \;=\; 0\]
If we define this quantity as $\tilde{J}^\mu$, we can write the above equation as
\[
\partial_\mu \tilde{J}^\mu = 0
\]
As discussed in section 5.1, there seems to be some association between the ideas that physics remain unchanged if you perform a global phase transformation on the mathematical object that describes state of a quantum system, and that probability is conserved. It was not a coincidence, Noether's theorem makes it precise here, if we interpret $\bar\psi\gamma^\mu\psi$ as the probability 4-current.
\[
\frac{\partial \tilde{J}^0}{\partial t} \;+\; \nabla \cdot \vec{\tilde{J}} \;=\; 0
\]
If we multiply this probability 4-current $\tilde{J}^\mu$ with the charge $q$, we get the familiar 4-current from electromagnetism $J^\mu$. Hence we can write the lagrangian \eqref{QEDlagrangian} as follows:

\[\mathcal{L} \;=\; \mathcal{L}_{\text{kinetic for matter}}
\;-\; J^\mu(x)\,A_\mu(x)
\;-\; \frac{1}{4\mu_0}\,F^{\mu\nu}F_{\mu\nu}
\]
If you recall, this is precisely the Lagrangian \eqref{eq: ED_Lagrangian}, we had built in section 4 for \textbf{Electrodynamics}. But, this time we have done it \textbf{only} from the principles of symmetry and local U(1) invariance. 
\paragraph{Remark.} The result we have obtained captures the essence of what it means for a theory to be a \textit{gauge theory}. The electromagnetic interaction does not need to be added artificially into the Dirac Lagrangian, but rather it emerges as the minimal and unavoidable consequence of imposing local $U(1)$ symmetry. This same gauge principle underlies the full Standard Model of particle physics: strong nuclear interaction arises from local $SU(3)$ symmetry, and the weak interaction arises from local $SU(2)$ symmetry (after electroweak symmetry breaking). Thus, by understanding QED through the lens of gauge invariance, we have uncovered the conceptual foundation shared by all forces described in the Standard Model Lagrangian.\\
While the detailed structure of $SU(2)$ and $SU(3)$ gauge theories lies beyond our present scope, the essential idea is now in place, and the reader is prepared to explore the full Standard Model (see \cite{griffiths2008} for a comprehensive introduction to particle physics).

\bibliographystyle{unsrt}
\bibliography{refs}

@book{schwichtenberg2018,
  author    = {Jakob Schwichtenberg},
  title     = {Physics from Symmetry},
  series    = {Undergraduate Lecture Notes in Physics},
  publisher = {Springer},
  address   = {Cham, Switzerland},
  year      = {2018},
  isbn      = {978-3-319-66630-3},
  doi       = {10.1007/978-3-319-66631-0}
}

@book{griffiths2017,
  author    = {David J. Griffiths},
  title     = {Introduction to Electrodynamics},
  edition   = {4th},
  publisher = {Cambridge University Press},
  address   = {Cambridge, UK},
  year      = {2017},
  isbn      = {978-1-108-42041-9},
  doi       = {10.1017/9781108333511}
}

@book{griffiths2008,
  author    = {David J. Griffiths},
  title     = {Introduction to Elementary Particles},
  edition   = {2nd revised},
  publisher = {Wiley-VCH},
  address   = {Weinheim},
  year      = {2008},
  isbn      = {978-3-527-40601-2},
  doi       = {10.1002/9783527618460}
}

\end{document}